\documentclass[
	reprint,
	amsmath,amssymb,
	aps,
]{revtex4-2}
\usepackage{graphicx}
\usepackage{accents}

\newcommand{\bcdot}[0]{\,\raisebox{0.1em}{\scalebox{0.6}{$\bullet$}}\;}
\newcommand{\pvec}[1]{{\vec{#1} \,}'}

\begin{document}

\title{Simulation of Refraction Through Voxel Volumes}
\author{Rainier Mobley$^{1}$}
\author{Jean Michel Létang$^2$}
\author{Max Langer$^1$}
\affiliation{%
	$^1$ Univ. Grenoble Alpes, CNRS,
	UMR 5225 VetAgro Sup, Grenoble INP, TIMC,
	Grenoble F-38000, France\\
	$^2$ INSA Lyon, Université Claude Bernard Lyon 1,
	CNRS, Inserm, CREATIS UMR 5220 U1294,
	Lyon F-69373, France
}%
\date{\today}

\begin{abstract}
Ray tracing in X-ray science
has typically been limited to descriptions
where refraction occurs only at interfaces,
ignoring the refraction occurring
due to the continuous variations in index of refraction
in the bulk of a material,
as are common in biological samples like bone.
We propose a method
to simulate X-ray images directly
on voxel grids using curved rays.
We first validate this method
compared to a model with an analytical solution,
and then investigate the effects of voxel size,
interpolation choice,
and illumination to grid alignment
on a few simulated test objects.
From these investigations,
we make several recommendations
to guide simulations using these techniques.
\end{abstract}

\maketitle

\section{Introduction}
As phase contrast imaging is growing in prominence,
understanding the effects of refraction
is becoming increasingly important in X-ray imaging.
There are a suite
of successful phase-constrast imaging techniques
such as
propagation-based imaging
\cite{snigirev1995possibilities},
analyser-based imaging
\cite{davis1995phase},
grating-based imaging
\cite{pfeiffer2006phase},
and
speckle-based imaging
\cite{morgan2012x,berujon2015near}
that measure images formed by the effects
of refraction and diffraction.
A thorough understanding of the image-forming process
is crucial to accurately transforming these images
into useful information
of the index of refraction of the target objects
and thus the electron density.

The data from these techniques are typically interpreted
using image-forming models based on the paraxial approximation
where all angular deviations of the coherently scattered light
are assumed to be small
\cite{paganin2006book}.
This leads to interpreting images
as various derivatives of a 2D straight-line projection
through the 3D index-of-refraction volume of the object.
The real part of the index of refraction of most materials
at most X-ray energies is very close to unity,
often differing on the order of one part per million,
so the approximation is highly accurate in many applications,
hence the great success of these methods.
However, it has been shown that 
for high-spatial-resolution detectors
and relatively thick objects
the approximation begins to fail,
even at X-ray energies
\cite{jadick2025accuracy}.
Simulations that take non-paraxial refractive effects
into account are thus important to develop.

Accurate simulations are valuable tools
for experimental design and algorithm development.
They can help plan experiments
by aiding to predict and optimize contrast relative to dose,
potentially saving crucial time that would otherwise
be spent tuning these parameters on site.
In addition, they can predict when the paraxial approximation is valid,
and simulate data sets that can avoid the ``inverse crime''
\cite{colton1998book}
of using the same simplified image-forming model to simulate synthetic data
as is assumed in an inverse algorithm
like any of the various flavors of X-ray computed tomography.

In addition to refraction,
there are three additional effects
that should be reproduced by an ideal simulation:
absorption, incoherent scattering,
and interference effects (diffraction).
The current suite of
state-of-the-art X-ray simulations strategies
handle various subsets of these effects.
These simulations fall in two camps:
wave mechanics and ray tracing techniques.
Wave mechanics techniques such as the multislice method directly propagate
a discrete representation of the wavefront.
Interference and absorption are naturally described with these techniques
\cite{li2017multislice}
and they have be extended to account for incoherent scattering
in the electron microscopy \cite{wang1990dynamical}.
Ray-tracing techniques propagate individual ray paths
to reconstruct wavefronts in a Monte Carlo manner.
Ray tracing naturally describes absorption and incoherent scattering
\cite{sarrut2021advanced}
and can be extended to account for interference effects
\cite{keller2016geometrical,
kouyoumjian1974uniform,
peter2014combining,
cipiccia2014inclusion,
steinberg2025wave}.
In this work,
we focus on ray tracing
so that the methods developed
can later be easily implemented
in existing Monte Carlo particle software.

Refraction in X-ray ray-tracing techniques
has primarily been modeled
as occurring only at interfaces between two materials
with different indices of refraction.
To simulate refraction through a given target object,
these interfaces typically are represented
with analytic geometric descriptions or with triangle meshes
\cite{langer2020towards}.
Such representations are well-suited for engineered objects
whose surfaces are already defined in terms of these representations;
however, 3D reconstructions of imaged objects
are usually represented in terms of voxel grids.
A method has been developed to refract
through implicit interface geometry
inferred directly from segmented voxel representations,
that is, voxels labeled as containing one of a short list of materials
\cite{tran2020modeling}.
These methods all assume that the index of refraction
of the underlying object is piecewise constant.
The rays thus propagate along straight lines between 2D interfaces.
Real objects, however, especially biological samples,
can have continuous variations
in the index of refraction throughout their volume.
These variations also cause refraction
that is neglected by state-of-the-art X-ray simulations.
To simulate this refraction with ray tracing requires rays
that curve as a function of the local index of refraction
as governed by the ray equation
\cite{sommerfeld1911anwendung}.
3D index-of-refraction volumes can be represented
with analytic geometric descriptions,
with tetrahedra meshes,
or with voxel grids.
We focus in this work on voxel grids
due to their widespread use.

Ray tracing methods with curved rays have been developed
in many other scientific domains including
geophysics
\cite{pereyra2000ray},
ultrasound
\cite{denis1995ultrasonic},
oceanography
\cite{halsne2023ocean,georges1986simulating},
atmospheric science
\cite{marks1995three,virieux2004paraxial,hobiger2008fast},
and
computer graphics
\cite{ihrke2007eikonal,cao2010interactive}.
The various
acoustic, gravity, radio, or visible light waves
in these fields
are governed by analogous wave mechanics,
and thus methods developed in these fields
can be an inspiration for new X-ray methods.

In this work,
we propose a method
to simulate X-ray images
directly on voxel grids,
allowing for curved rays.
We validate our technique by comparing to
a well-known historical analytic solution,
the brachistochrone problem.
We then investigate the effects of representing
interfaces between two materials with voxels.
We investigate the effects of the choice of interpolation
on a few simple simulated test objects.
Finally, to demonstrate the use of the method for imaging,
we simulate images of representative synthetic objects.

\section{Theory}
The regime
of geometrical optics describes
the propagation of light
in terms of rays,
that is, paths that are governed
by the ray equation,
in order to approximate
the underlying wave mechanics
\cite{lin2017advanced}.
It neglects effects
such as diffraction and interference,
and is a good approximation
in the case that the wavelength
is small compared to the structures
with which the light interacts.
The ray equation can be expressed as
	\begin{equation}
	\left( \vec{\nabla} n(\vec{r} \, ) \, \bcdot \, \frac{d\vec{r}}{ds} \right)
	\frac{d\vec{r}}{ds}
	+ n(\vec{r} \, ) \, \frac{d^2 \vec{r}}{ds^2}
	- \vec{\nabla} n(\vec{r} \, )
	= 0
	\label{ray_equation_main}
	\end{equation}
where $n$ is the index of refraction,
and $\vec{r}(s)$ is the ray path
parameterized by its arclength, $s$.
This equation
is derived from first principles
in appendix \ref{appendix_derivation}.
It is a non-linear partial differential equation.
There are some index-of-refraction volumes
that admit an analytic solution,
but in general it must be solved numerically.
Notably,
for a constant index of refection,
$\vec{\nabla} n = 0$,
the equation states that paths
have zero curvature,
$\frac{d^2 \vec{r}}{ds^2} = 0$,
thus the rays are straight
and continue in their initial direction.
In general, a gradually changing
index of refraction results in curved paths.
A discontinuous step in index of refraction
results in an undefined $\vec{\nabla} n$
and cannot be treated by the ray equation alone.

To numerically solve the ray equation,
we have implemented
a simple, first-order integrator,
that is,
	\begin{equation}
	\vec{r}_{j + 1} = \vec{r}_j + \Delta s \, \vec{v}_j
	\label{integrator1}
	\end{equation}
	\begin{equation}
	\vec{v}_{j + 1} = \vec{v}_j
	+ \Delta s \, \frac{
		\vec{\nabla} n
		- \left( \vec{\nabla} n \bcdot \vec{v}_j \right) \, \vec{v}_j
	}{n}
	\label{integrator2}
	\end{equation}
where $\vec{r}_j$ is the $j^\text{th}$ position,
$\vec{v}_j$ is the $j^\text{th}$ value
of $\frac{d\vec{r}}{ds}$,
and $\Delta s$ is a constant step size.
There are more sophisticated options,
but we chose this simple, first-order integrator
to be easily compatible
with Monte Carlo particle transport
in future work.
With this integrator,
an index-of-refraction volume
defined over all space,
and a sufficiently small step size,
we can solve the path of a ray
given any initial position and direction.

A length scale
to which we can compare
our step size
is that of the radius
of curvature of a ray.
We can use the ray equation,
Eq.~\ref{ray_equation_main},
to calculate
the radius of curvature as
	\begin{equation}
	R_\text{curvature}
	= \frac{n}{|\vec{\nabla} n| \sin(\theta)}
	\; ,
	\label{radius_of_curvature}
	\end{equation}
where $\theta$ is the angle between 
the ray and $\vec{\nabla} n$.
We can also define
a point-wise minimum
radius of curvature as
	\begin{equation}
	R_\text{tightest curvature}
	= \frac{n}{|\vec{\nabla} n|}
	\label{radius_of_curvature_tighest}
	\end{equation}
which calculates
the smallest radius
of curvature a ray can have
at a given position independent
of the particular path.


If the index-of-refraction volume
is represented in terms
of values on a voxel grid,
$n_{ijk}$, rather than
an analytic expression,
we must to choose
how to interpolate at positions
between the voxel grid points
so that we can define
an analytic expression
for the index-of-refraction volume
at any position,
$n(\vec{r} \, )$.
Assuming the constraint
that the value
of the index of refraction
must be equal
to a given voxel value
at the center of that voxel,
that is,
	\begin{equation}
	n(\vec{r}_{ijk} \, ) = n_{ijk}
	\; ,
	\end{equation}
where $\vec{r}_{ijk}$
is the position
of the ${ijk}^\text{th}$
voxel center,
there remain arbitrarily many ways
to interpolate between
the given values
at positions between voxel centers,
each with its own set
of advantages and inconveniences.
Each style of interpolation corresponds
to a choice a basis
with the given voxel values as coefficients,
and this choice decides
in which space our
index-of-refraction volumes are constrained.
That is,
	\begin{equation}
	n(\vec{r} \, ) =
	\sum_{ijk} n_{ijk} \, \xi_{ijk}(\vec{r} \, )
	\; ,
	\end{equation}
where $n_{ijk}$ is the ${ijk}^\text{th}$ voxel value,
and $\xi_{ijk}(\vec{r} \, )$
is the ${ijk}^\text{th}$ basis function.
To respect our voxel-value constraint,
we have
	\begin{equation}
	\xi_{ijk}(\vec{r}_{i'j'k'} \, ) =
	\delta_{ii'} \,
	\delta_{jj'} \,
	\delta_{kk'} \,
	\; ,
	\end{equation}
where the $\delta_{\alpha \alpha'}$
are Kronecker deltas.
That is, the ${ijk}^\text{th}$ basis function
must be 1 at the ${ijk}^\text{th}$ voxel center
and 0 at the centers of every other voxel.
We will also assume
	\begin{equation}
	\xi_{ijk}(\vec{r} \, ) =
	\xi_{000}(\vec{r} - \vec{r}_{ijk} + \vec{r}_{000})
	\; ,
	\end{equation}
that is,
each of the basis functions
is the same shape translated
to be centered on its
respective voxel grid point.
Knowing the basis functions also
allows us to calculate
the gradient of the index of refraction
at any position as
	\begin{equation}
	\vec{\nabla} n(\vec{r} \, ) =
	\sum_{ijk} n_{ijk} \, \vec{\nabla} \xi_{ijk}(\vec{r} \, )
	\; .
	\end{equation}

\begin{figure}[ht]
	\centering
	\includegraphics[width=0.5\textwidth]{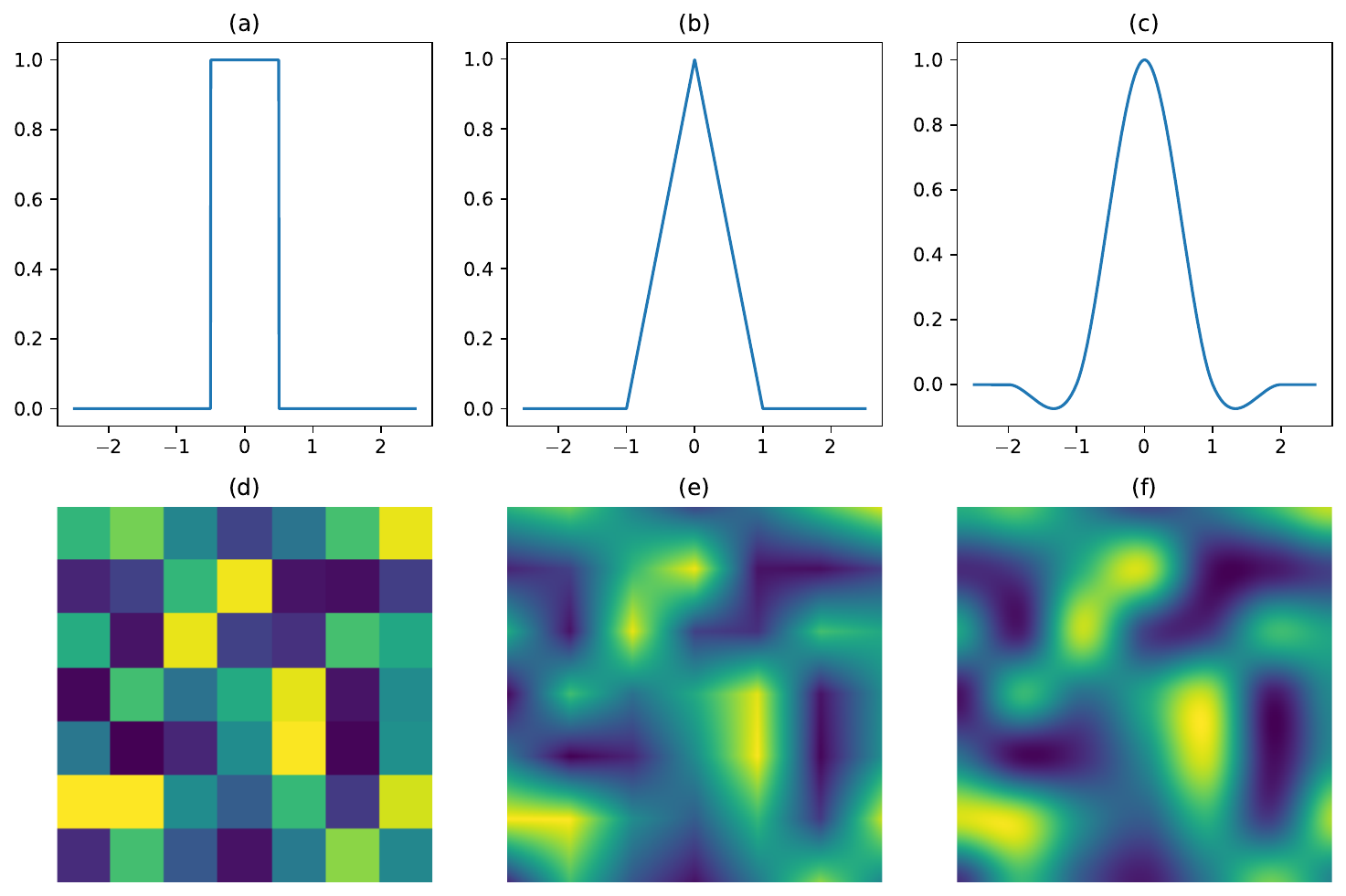}
	\caption{
	(a - c) three simple choices
	of 1D interpolation basis functions:
	(a) nearest neighbor,
	(b) linear interpolation,
	and (c) cubic interpolation.
	(e - g) examples
	of (e) 2D nearest neighbor, (f) bilinear,
	and (g) bicubic interpolation
	applied to the same grid
	of uniform random samples.
	}
	\label{fig_interpolation_both}
	\end{figure}

The simplest choice
of interpolation is that
of nearest neighbor interpolation.
That is, outputting
the voxel value
of the closest voxel grid point.
The shape of its basis function
in 1D is illustrated in
Fig.~\ref{fig_interpolation_both}a,
and a 2D example
of the interpolation
is shown in
Fig.~\ref{fig_interpolation_both}d.
This interpolation
is highly efficient in that
we only need to know
a single voxel value
to compute it
for a given position.
It has the problem, though,
that the gradient is zero
except at the boundaries
between voxels at
which there
are discontinuous jumps.
Eq.~\ref{ray_equation_main}
is thus inapplicable.
Refraction in this case
of discontinuous interfaces
can be calculated
with the Fresnel equations,
but refraction
from a regular lattice
of fictitious interfaces
will be full
of detrimental artifacts.
For example,
forward propagating
rays parallel
to the lattice
will not refract.
We thus did not explore
this interpolation choice.

The second simplest
interpolation choice
is that of
trilinear interpolation.
A detailed description
of how
it is calculated
is presented in appendix
\ref{trilinear_appendix}.
This method requires the 8
neighboring voxel values
of a given position
to calculate,
so it is slower than nearest neighbor
but still quite efficient.
An illustration of its
basis function shape in 1D
and an example 2D interpolation
are shown in
figures \ref{fig_interpolation_both}
(b) and (e),
respectively.
This choice
of interpolation
defines a gradient
in the volume between voxels,
so we can calculate rays with
Eq.~\ref{ray_equation_main}.

The next higher-order interpolation
is that of tricubic interpolation.
A detailed description
of how it is calculated
is presented in appendix
\ref{tricubic_appendix}.
This method requires
64 voxel values
to compute an interpolation
at a single point,
not only the neighboring voxel values
but the neighbors' neighbors.
An illustration of its
basis function shape in 1D
and an example 2D interpolation
are shown in figures \ref{fig_interpolation_both}
(c) and (f),
respectively.
It defines a continuous gradient field
everywhere in our refraction volume,
so the rays it yields with
Eq.~\ref{ray_equation_main}
should be well-behaved.


\section{Numerical Experiments}

\subsection{Brachistochrone Problem}
\label{brachistochrone_section}
The brachistochrone problem
is the problem in classical mechanics
of finding the curve of fastest descent
that connects two points.
We can think of this curve
as the shape of a wire
threaded through
a very slippery bead.
The question becomes,
what shape should
we bend this wire
to have the bead slide
the fastest
between a given start and end point
when acted upon only by gravity?
Bernoulli famously solved the problem
by reinterpreting it
as the path of a ray of light
in a medium whose index of refraction
varies as a function of height.
This is the interpretation
we will consider here
to test our technique compared
to a known analytical solution
to the ray equation.

The potential energy
of the bead as a function
of its height, $z$, is
	\begin{equation}
	U(z) = m \, g \, z
	\,
	\end{equation}
where $m$ is the mass
and $g$ is the acceleration
due to gravity.
The kinetic energy is
	\begin{equation}
	K(v) = \frac{1}{2} \, m \, v^2
	\,
	\end{equation}
where v
is the speed of the bead.
The total energy is 
	\begin{equation}
	E = U(z) + K(v)
	\end{equation}
which is constant.
Assuming the bead starts
with v=0 and z=0, and thus E=0,
we can write the speed
as a function of height.
That is,
	\begin{equation}
	v(z) = \sqrt{-2 \, g \, z}
	\; ,
	\end{equation}
where z is a negative number.

The key ideas to interpreting
the problem in terms
of the ray equation are
that the index of refraction
is the ratio
of the speed of light
in vacuum, $c$,
to the speed of light
in a medium,
that is,
	\begin{equation}
	n = \frac{c}{v}
	\; ,
	\end{equation}
and Fermat's principle
that light rays take
the path of least travel time
between two points.
We can thus choose a $n(z)$
to reproduce a ray
that follows the same
fastest descent trajectory
as the classical bead.
We can see from
Eq.~\ref{ray_equation_main}
that any proportionality
doesn't effect the path
(only the phase),
so we can imagine
a medium that has
a index of refraction
varying as
	\begin{equation}
	n(z) = \frac{1}{\sqrt{-z}}
	\end{equation}
to reproduce
the brachistochrone path.

\begin{figure}[ht]
	\centering
	\includegraphics[width=0.49\textwidth]{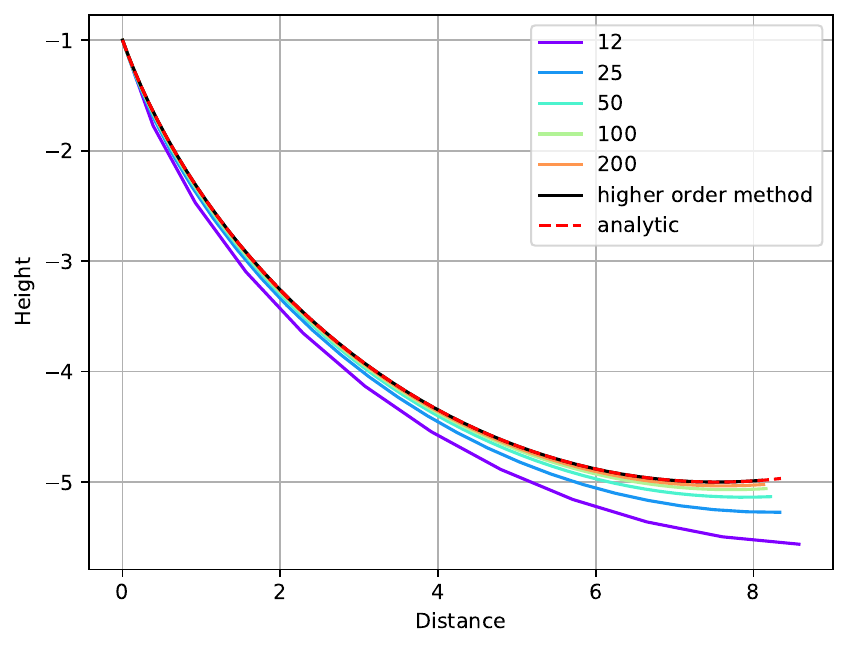}
	\caption{
		Comparison
		of several numerical solutions
		and the analytic solution
		of the ray-equation formulation
		of the brachistochone problem.
		The first-order solutions
		are labeled by the number
		of steps used.
		The minimum radius of curvature
		of the analytic solution
		is 4.47 units.
		The total arclength
		of each of the numerical solutions
		is 9.5 units.
		For the 200 step solution,
		the minimum radius of curvature
		is about 100 times the step size.
		The path errors are
		12.3\%,
		5.6\%,
		2.7\%,
		1.4\%,
		and 0.7\%
		for the 12, 25, 50, 100, and 200
		step solutions respectively.
		The path error
		for the higher-order method
		is 0.01\%.
		%
	}
	\label{brachistochrone_fig}
	\end{figure}

Fig.~\ref{brachistochrone_fig}
shows our implementation
of the first-order
integrator in
Eq.~\ref{integrator1} and Eq.~\ref{integrator2}
at various step sizes
compared to the analytic solution
and a higher-order
numerical solution.
The reported path-error
metric is defined in equation
\ref{path_error}
of appendix
\ref{path_error_section}.
These results suggest
that step size about 1\%
the minimum radius of curvature
results in an acceptable
level of path error,
which we will use
as a heuristic
in our later explorations.

\subsection{Voxel Interfaces}
\label{interfaces_section}
Discontinuous jumps
in index of refraction
at interfaces
between materials
cause light
to refract or reflect
at the interfaces according
to the Fresnel equations.
We will call
these discontinuous interfaces
with infintesimal thickness
\textbf{\textit{sharp}} interfaces.
The ray equation
as it is formulated in
Eq.~\ref{ray_equation_main},
cannot calculate
the behavior of rays
at sharp interfaces
due its reliance
on the gradient
of the index of refraction,
which is not defined
at these interfaces.
For an index of refraction volume
defined everywhere
with both sharp interfaces
and variations in the bulk
between the interfaces,
the appropriate strategy
to simulate rays
is to use the ray equation
to calculate the curved rays
in the bulk
and to use
the Fresnel equations
at sharp interfaces.
The problem
is that,
when given only
a voxel representation
of an index-of-refraction volume,
any sharp interfaces
must be inferred and defined
from the voxel values alone.
Algorithms for doing so
have been devised
for the constrained case
of piecewise-constant
refraction volumes 
\cite{tran2020modeling},
but none have
been developed
for the general case.
These algorithms
are left for future work.

\begin{figure}[ht]
	\centering
	\includegraphics[width=0.49\textwidth]{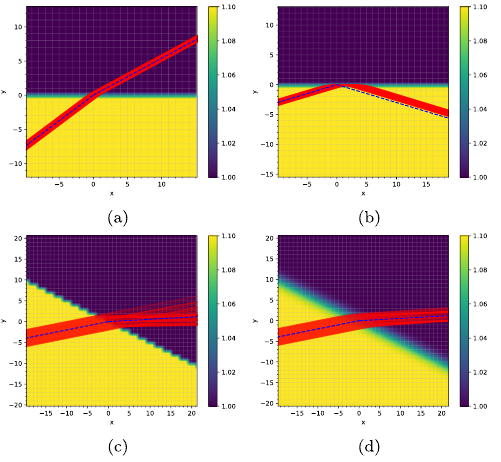}
	\caption{
		Rays refracted
		through various pixelated
		linear interfaces
		interpolated bilinearly.
		The red lines
		are the propagated rays.
		The dotted lines are the rays
		one would expect
		using the Fresnel equations
		across a perfectly sharp step
		with the same magnitude.
		The pixels
		have an edge length of 1 unit.
		(a)
		refraction through
		a linearly interpolated step
		aligned with the pixel grid.
		(b) reflection
		from the same aligned interface.
		(c) refraction through
		a naively pixelated interface
		skewed at an angle
		relative to the pixel grid.
		(d)
		refraction through
		a band-limited version
		of the same skewed interface.
		The band-limited interface
		was constructed
		by convolving a sharp interface
		by a Gaussian,
		that is,
		applying a Gaussian filter
		to the frequency components,
		before sampling on a grid.
		The Gaussian was chosen
		to have a standard deviation
		of $4 / \pi$ distance units.
		This corresponds
		to the spacial sampling frequency
		being 8 times
		the standard deviation
		of the Gaussian
		in frequency space.
		The minimum
		radius of curvature
		(eq.
		\ref{radius_of_curvature_tighest})
		for (a) and (b)
		is 10 units.
		For (c) and (d)
		it is 7 and 11 units,
		respectively.
		The rays were calculated
		with a step size
		of 0.025 units.
	}
	\label{fig_interface}
	\end{figure}

Neglecting the possibility
of sharp interfaces entirely
and simply interpolating
the voxel values defines
an index-of-refraction volume
with only
continuous bulk variation,
even between adjacent voxels
with very large jumps
in index of refraction.
When interpolated,
these large jumps between voxels
create thin regions
of large gradient
in the refractive index.
We'll call
thin but continuous
regions of transition
between two
indices of refraction
\textbf{\textit{soft}} interfaces.
The results of applying
the ray equation
to various
voxel-representation
soft interfaces
are presented
in Fig.~\ref{fig_interface}.

From the results in
Fig.~\ref{fig_interface} (a)
and Fig.~\ref{fig_interface} (b)
we can see that,
when the soft interface
is aligned
with the voxel grid,
the refraction
through it
bends rays to exit
with the same angle
that the Fresnel equations
would predict
for a sharp interface.
In the case
of total internal reflection,
the rays receive a shift
reminiscent of the 
Goos-Hänchen effect
\cite{goos1947neuer},
but at the scale
of the voxel size
rather than the wavelength.
The effect
of partial internal reflection
is lost.
If we take an underlying
sharp interface
that is misaligned
with the voxel grid
and simply sample
it at the voxel positions,
we define a jagged
soft interface
that scatters rays
in many directions,
as seen in Fig.~\ref{fig_interface} (c).
This is an aliasing problem.
A sharp interface
contains much too high
of frequency information 
to be accurately represented
by a voxel grid
according to
the Shannon-Nyquist theorem.
Filtering out
the high-frequency components
by convolving
the underlying
step function
with a Gaussian
softens the interface.
For any given voxel size
and desired accuracy,
there is
a sufficiently large
Gaussian blur
such that
the voxel values
carry enough information
to adequately reconstruct
the underlying soft interface
with Shannon interpolation.
The result in
Fig.~\ref{fig_interface} (d)
uses bilinear interpolation,
causing
slight artifacts
in the refracted beam,
but significantly
less than in
Fig.~\ref{fig_interface} (c).
Further increasing
the width of the smoothing
further reduces
the angular deviations.
The following investigations
focus only on bulk refraction
in interface-free objects.

\subsection{Interpolation Comparison}
\label{section_interpolation_comparison}
In this section, we apply
our voxel technique
to various interface-free 3D objects
to simulate images
under plane-wave illumination.
These images are simulated
by counting the number of rays
incident on each pixel,
as explained in appendix
\ref{appendix_amplitude}.
The initial positions of the rays
are sampled uniformly on
a square surface that is larger
than the object or detector.
Their initial directions
are perpendicular to this surface
towards the object.

\subsubsection{Gaussian Target}
\label{section_gaussian}
\begin{figure}[!t]
	\centering
	\includegraphics[width=0.5\textwidth, trim= 0 20mm 0mm 24mm, clip]{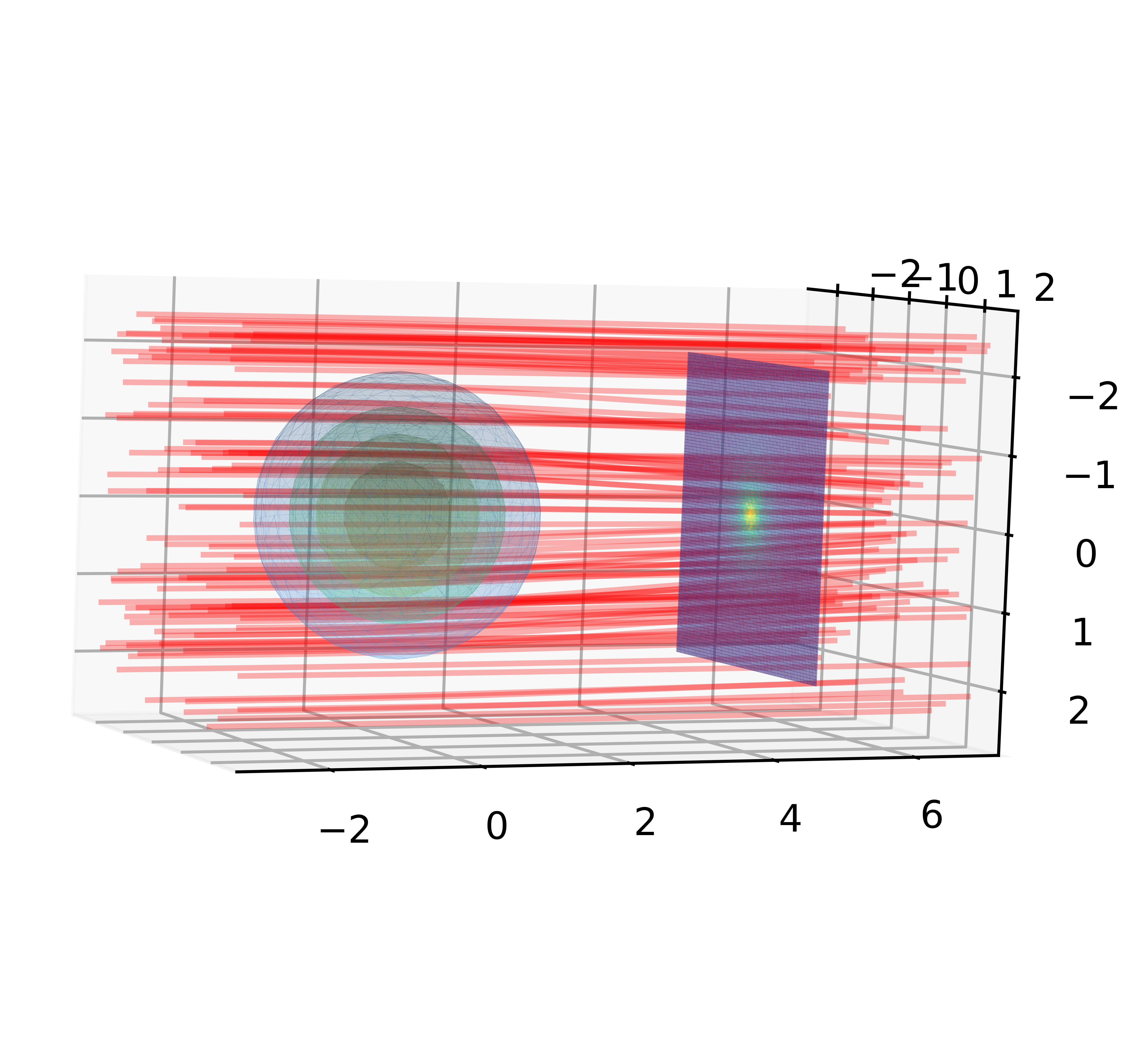}
	\caption{
		3D view
		of the ray tracing setup
		for a Gaussian target.
		The concentric spheres
		are isosurfaces
		of a Gaussian
		centered at the origin.
		The width of the Gaussian,
		by this we are referring
		to twice its standard deviation,
		is 2 units
		and the amplitude is 0.05
		above a baseline of 1.
		The red lines
		are a random subset
		of the traced rays
		originating from the left.	
		The simulated detector
		is 4 units by 4 units
		and is positioned
		5 units away.
	}
	\label{fig_analytic_gaussian_3D}
	\end{figure}

\begin{figure}[!t]
	\centering
	\includegraphics[width=0.50\textwidth]{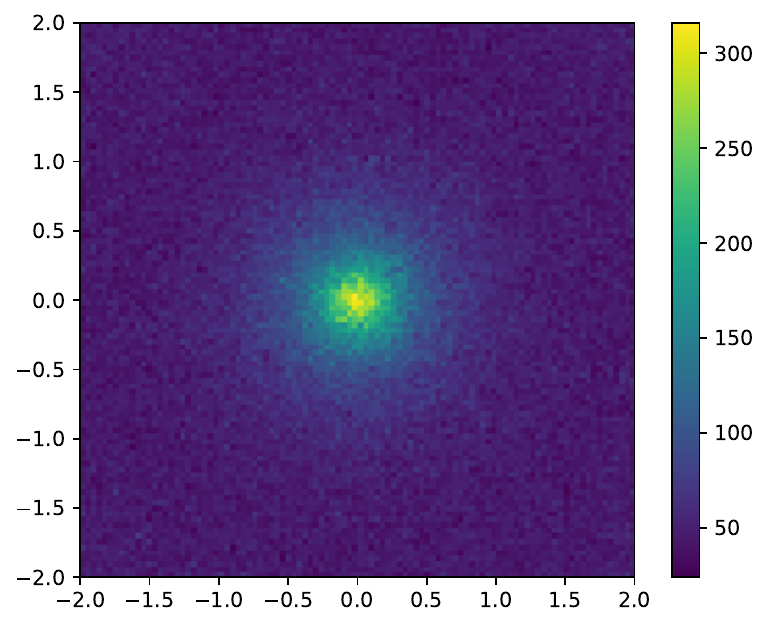}
	\caption{
		Simulated measurement
		for and analytic Gaussian
		corresponding to the detector
		in Fig.~\ref{fig_analytic_gaussian_3D}.
		This was calculated
		by propagating 800,000 rays
		and counting the number
		that intersect each pixel
		to create a image.
		The detector is
		100$\times$100 pixels.
		The color scale corresponds
		to ray counts per pixel.
	}
	\label{fig_analytic_gaussian}
	\end{figure}

The first
simulated test object
is that of a 3D Gaussian.
The imaging setup
is shown in
Fig.~\ref{fig_analytic_gaussian_3D}.
To create a ground truth
with which to compare
our voxel results,
we simulated an image
by refracting rays through
the purely analytic (no voxels)
representation of the Gaussian.
The resulting image
is shown in
Fig.~\ref{fig_analytic_gaussian}.
For the voxel results,
the Gaussian was represented
with trilinear and tricubic 
voxel grids
at a variety of resolutions.
For each, an image was simulated
using rays with the same initial conditions
as in the ground truth image.

	\begin{figure*}[t]
	\centering
	\includegraphics[width=\textwidth]{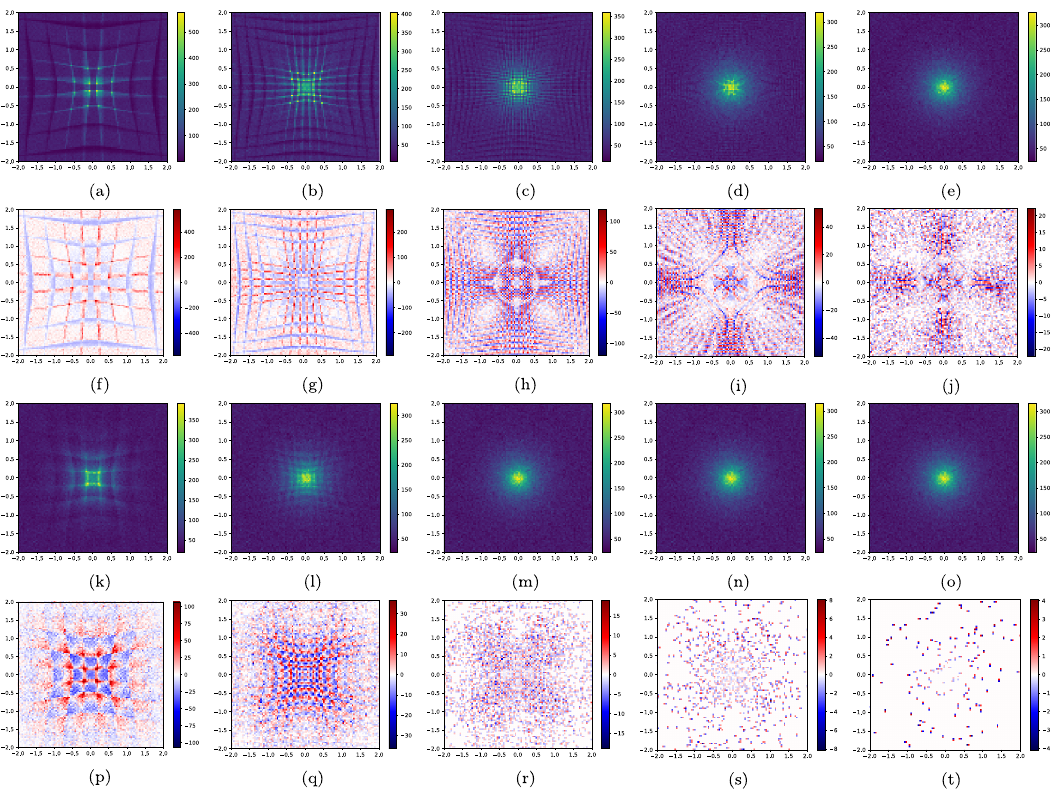}
	\caption{
		The results
		of propagating 800,000 rays
		through different voxelizations
		of the same underlying Gaussian
		and imaging setup as
		described in Fig.~\ref{fig_analytic_gaussian_3D}
		(a - e)
		the results
		for voxels that are
		$\frac{1}{4}$,
		$\frac{1}{8}$,
		$\frac{1}{16}$,
		$\frac{1}{32}$,
		and $\frac{1}{64}$
		the width of the Gaussian,
		interpolated trilinearly.
		The color scales corrspond
		to ray counts per pixel.
		(k - o)
		the results
		with the same sequence of voxel sizes
		interpolated tricubically.
		(f - j)
		the percent differences
		of the trilinear calculations above
		relative to the analytic calculation
		shown in Fig.~\ref{fig_analytic_gaussian}.
		The rays in both calculations
		have the same initial conditions.
		The normalized
		root mean square errors,
		Eq.~\ref{rms_error},
		are
		51.4\%,
		40.1\%,
		25.2\%,
		9.6\%,
		and 4.1\%
		respectively.
		Similarly, (p - t) depict
		the percent differences
		between tricubic calculations
		and the analytic calculation.
		The errors are
		21.1\%,
		8.2\%,
		2.4\%,
		0.7\%,
		and 0.3\%
		respectively.
		In (t), the differences
		consist of a small number
		of individual rays
		that are only one pixel off
		from the analytic case.
	}
	\label{fig_voxel_gaussian}
	\end{figure*}
The results are pictured in
Fig.~\ref{fig_voxel_gaussian}.
Artifacts arising
from the voxel representations
are visible in the images,
but lessen progressively
for the higher resolution
voxel grids.
The discontinuous gradients inherent
in the trilinear representations
cause the images to be discontinuous
as well.
The resulting artifacts
become less apparent
in the higher-voxel-resolution images
because the voxel-scale artifacts
are averaged within a single pixel.
At intermediate resolution,
the artifacts manifest
as Moiré patterns.
The images
of the tricubic representations
are continuous
and the artifacts
disappear in an absolute sense
as the voxel resolution increases.

\begin{figure*}[t]
	\centering
	\includegraphics[width=\textwidth]{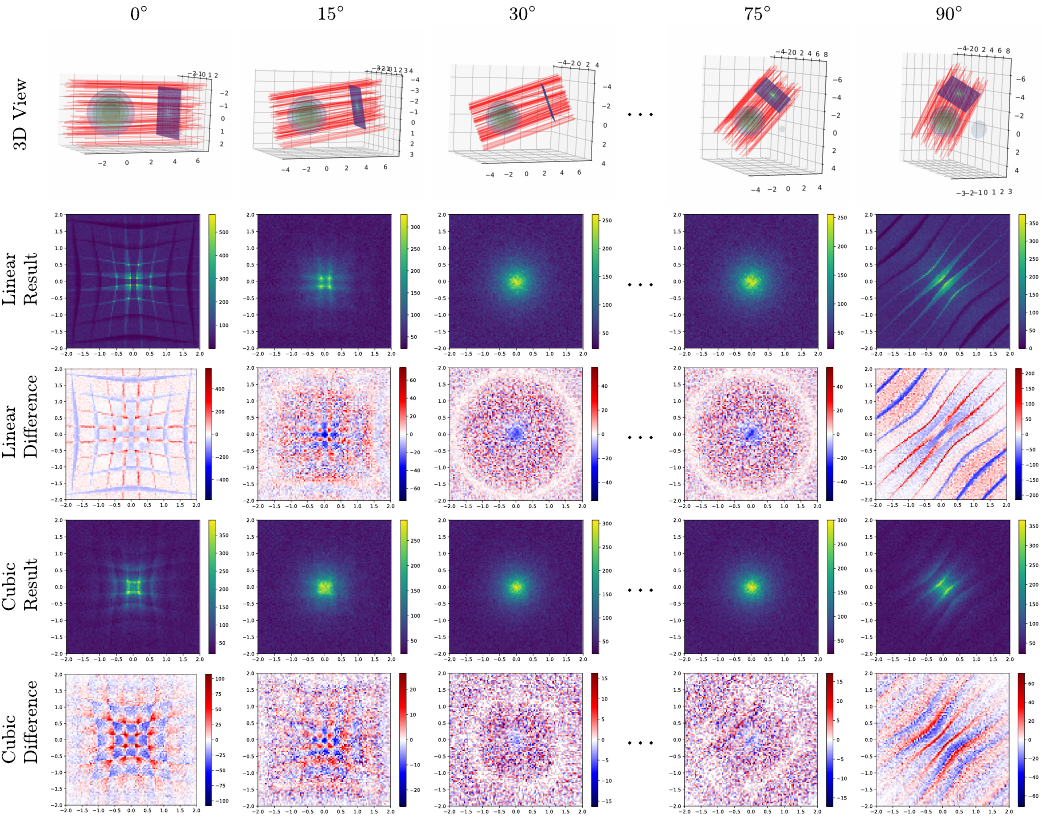}
	\caption{
		%
		The results
		of propagating 800,000 rays
		through the same
		voxelized Gaussian
		at different orientations.
		The detector and underlying Gaussian
		are the same as described
		in Fig.~\ref{fig_analytic_gaussian_3D}.
		The columns correspond
		to the results when
		the beam and detector
		are rotated by the associated angle
		about the [1, 1, 0] axis.
		The rows are as following:
		(3D View)
		the 3D views
		of the detector,
		the object,
		and a random subset of the rays
		at the specified orientations.
		(Linear Result)
		the result images
		when the object is
		interpolated tricubicly.
		(Linear Differences)
		the percent differences
		of the trilinear calculations
		relative to the analytic calculation
		shown in Fig.~\ref{fig_analytic_gaussian}.
		The normalized
		root mean square errors,
		Eq.~\ref{rms_error},
		are
		51.4\%,
		15.7\%,
		11.2\%,
		11.4\%,
		and 36.8\%
		respectively.
		(Cubic Differences)
		the percent differences
		between tricubic calculations
		and the analytic calculation.
		The errors are
		21.1\%,
		6.1\%,
		3.4\%,
		3.4\%,
		and 15.8\%
		respectively.
	}
	\label{rotation_fig}
	\end{figure*}

The particular illumination used
for the results in
Fig.~\ref{fig_voxel_gaussian},
a plane wave
aligned with the
voxel grid,
is the worst-case scenario
for voxel artifacts.
The affects decrease substantially
when the illumination angle
is far from aligned with the voxel grid.
This effect is demonstrated
in Fig.~\ref{rotation_fig},
for a coarsely voxelized Gaussian
illuminated at various angles.

\subsubsection{Structured Objects}
\label{section_lumpy}
\begin{figure*}[ht]
	\centering
	\includegraphics[width=\textwidth]{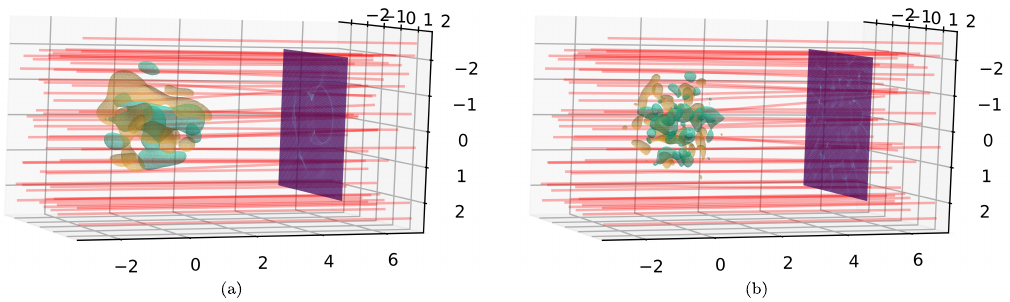}
	%
	\caption{
		3D views
		of the ray-tracing setup
		for two structured objects.
		Both objects
		are filtered noise
		modulated
		by a Gaussian envelope
		on top of a constant baseline
		The shapes plotted
		in teal and yellow
		are two isosurfaces
		of the object equidistant
		from the mean value,
		one above one below.
		The red lines
		are a random subset
		of the propagated rays.	
		The detectors
		each depict
		a simulated measurement.	
		The objects are generated
		from uniform random noise.
		For the object in (a),
		the noise is filtered
		with a hard frequency cutoff
		such that the average size
		of its features is 0.5 units.
		For the object in (b),
		the cutoff is such that
		the average size
		of its features is 0.25 units.
		The standard deviation
		of the filtered
		but unmodulated noise is 1.
		The height
		of the Gaussian modulation envelope
		is 0.0125.
		The background value is 1.
	}
	\label{fig_lumpy_3D_compound}
	\end{figure*}

The next simulated objects
are structured objects designed
to resemble inhomogeneous materials
with variations in refractive index
on some characteristic scale,
like bone.
The objects were generated
by modulating filtered noise
with a Gaussian envelope.
The noise is uniformly random,
ranging from -1 to 1.
The filter
is a hard cutoff
in frequency space
at the spatial frequency reciprocal
to the desired characteristic scale.
Two voxel volumes
were simulated,
each at a different
characteristic scale.
Refraction was simulated
through each voxel volume
using both
trilinear
and tricubic
interpolation.
The imaging setups are shown in 
Fig.~\ref{fig_lumpy_3D_compound}.

Note that to improve figure clarity, 
we intentionally amplified the magnitude 
of the variations of index of refraction
beyond physical limits reported for bone tissue 
in the X-ray regime (typical decrements are in 
the $[10^{-6},10^{-5}]$ range). 
This modeling choice affects only the absolute 
scale of contrast effects, 
not the underlying relationships
between structural parameters 
and observed signal modulations. 
Consequently, the insights derived from these simulations 
remain applicable to experimentally relevant conditions 
where the propagation distance would be 
much larger than the object size.

\begin{figure*}[ht]
	\centering
	\includegraphics[width=\textwidth]{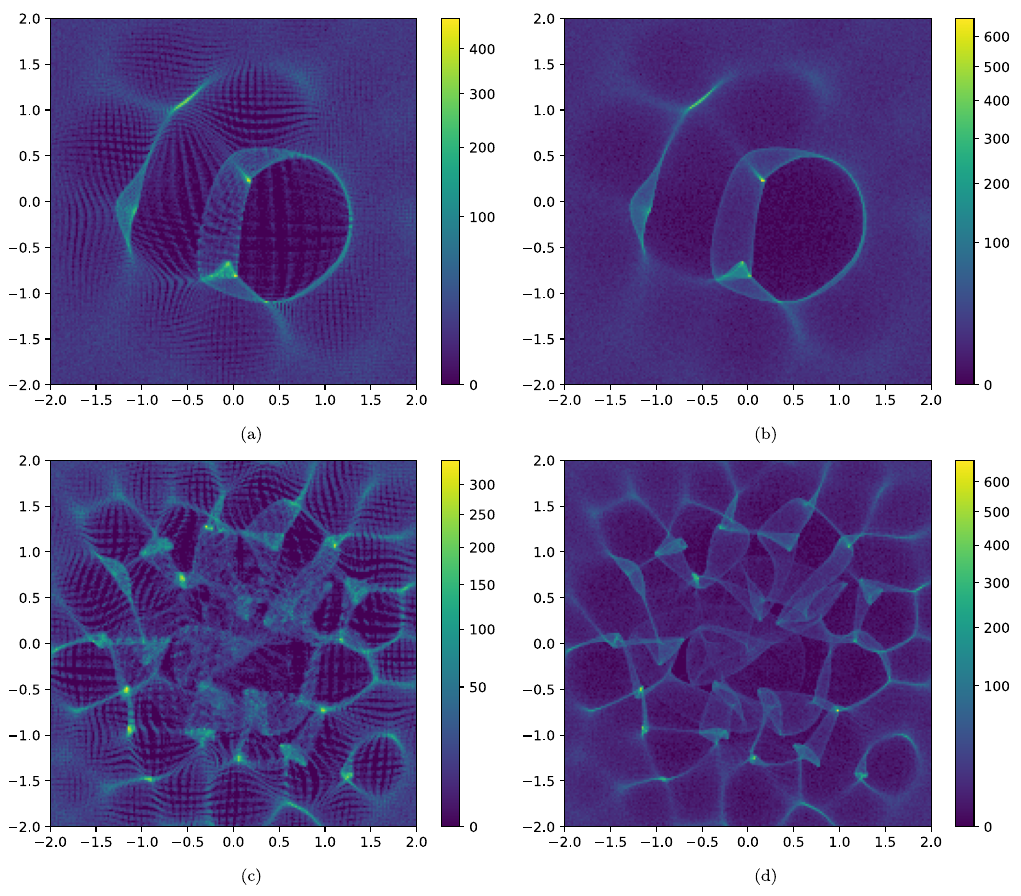}
	\caption{
		The images formed
		by the structured objects
		in Fig.~\ref{fig_lumpy_3D_compound}
		with the illumination aligned
		with the voxel grid.
		(a) and (b)
		images for the object
		with coarser features
		when interpolated
		trilinearly
		and tricubically,
		respectively.
		(c) and (d)
		measurements
		for the object
		with finer features
		interpolated trilinearly
		and tricubically,
		respectively.
		The detector is 200$\times$200 pixels
		and the images are displayed
		with a squareroot-transformation color scale.
	}
	\label{fig_lumpy_histogram_compound}
	\end{figure*}

\begin{figure*}[ht]
	\includegraphics[width=\textwidth]{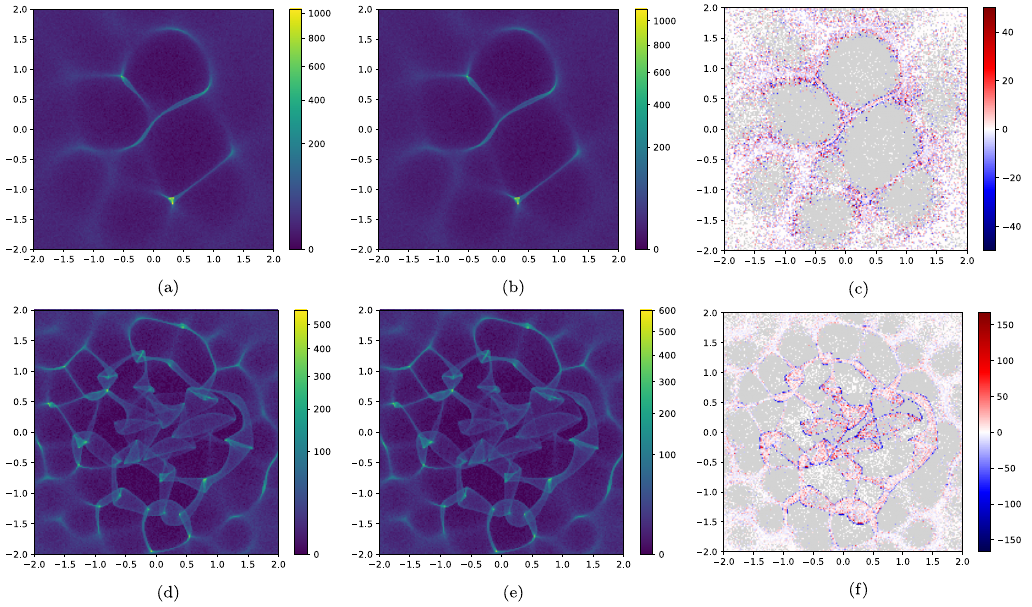}
	\caption{
		The images formed by the structured objects
		with the illumination misaligned
		from the voxel grid by $45^\circ$
		about the [1, 1, 0] axis.
		(a) and (b)
		the images for the object
		with coarser features
		when interpolated
		trilinearly
		and tricubically,
		respectively.
		(c) the difference between them
		with the tricubic subtracted from the trilinear
		expressed as a percentage of the tricubic image value.
		The regions
		with too few rays for good statistics
		are colored gray.
		(d) and (e)
		measurements
		for the object
		with finer features
		interpolated trilinearly
		and tricubically,
		respectively.
		(f) the difference between them as above.
	}
	\label{fig_lumpy_histogram_compound_rot}
	\end{figure*}

The images are shown in
Fig.~\ref{fig_lumpy_histogram_compound}.
Artifacts are clearly visible
in the tricubic interpolation result,
whereas the tricubic interpolation result
appears smooth.
The illumination in this case
was the worst-case scenario
of plane wave illumination
aligned with the voxel grid.
When the voxel grid
is rotated at an angle
relative to the illumination,
the obvious artifacts disappear
and are replaced with minor differences,
as seen in
Fig.~\ref{fig_lumpy_histogram_compound_rot}.

\section{Discussion}
%
%
%
%
%
%
Our initial validation
of the proposed technique
with an analytic solution
shown in
Fig.~\ref{brachistochrone_fig}
demonstrates that the presented technique
approaches the analytic solution
when using sufficiently small step sizes.
These results suggest
that a step size of about 1\%
the minimum radius of curvature
anywhere in the refraction volume
will yield rays with
an acceptable level of path error.
A possible improvement in future work could be
to use a dynamic step size
of about 1\% the minimum radius
of curvature in the local region.
This strategy can gain efficiency
by taking larger steps
in the regions where the rays
are straighter,
thus calculating less steps overall.
In addition, in Eq.~\ref{ray_equation},
the value of $|\frac{d\vec{r}}{ds}|$
is a conserved quantity as explained
in appendix \ref{appendix_ray_equation}
and should never deviate from unity.
This fact is not built
into this choice of integrator.
A more sophisticated choice
of an appropriate symplectic integrator
that respects this conserved value
would likely be more efficient.
There could also be significant gains
in employing a higher-order integrator.
These strategies would be
more complicated adapt to existing
Monte Carlo particle transport algorithms.

The results of our investigation
of refraction through interfaces
shown in Fig.~\ref{fig_interface}
demonstrate that replacing
a sharp interface,
a discontinuous interface
between to indices of refraction,
with a soft interface,
a thin but continuous
region of transition,
introduces artifacts
in the form of a displacement of the exit ray
at a scale on the order
of the width of the smooth interface.
The effect of partial internal reflection
is also lost.
Notably, though, the exit angles are preserved.
We also saw that describing soft interfaces
on a voxel grid sufficiently finer
than the width of the soft interface
introduces only small angular artifacts.
This suggests that refraction through
voxel representations
of sufficiently antialiased objects
can be well described
without the use of sharp interfaces,
and that the refraction
through these antialiased
voxelized objects
can reproduce important features
of the refraction
through the unfiltered
underlying object,
provided these features
are significantly larger
than the voxel size.
In the case of simulating refraction
from a CT reconstruction,
the reconstruction itself is naturally
an antialiased version of the true object,
so the simple ray propagation
with no consideration of sharp interfaces
presented here may be sufficient.
The exact criteria
for when this simple description
is satisfactory is left for future work.
A sophisticated simulation
would use bulk refraction,
and sharp interfaces together,
as well as absorption and
statistical scattering distributions
in the bulk and at sharp interfaces
that describe any sub-voxel
scattering processes.

The investigation in section
\ref{section_interpolation_comparison}
reveals several possible sources of voxelization
artifacts that should be
taken into account for simulations.
The results pictured in
Fig.~\ref{fig_voxel_gaussian}
demonstrate the effects of voxelization
in the case of plane-wave illumination
aligned with the voxel grid.
The most obvious effect is that
of the discontinuous blocky artifacts present
in the images calculated with trilinear interpolation.
They are caused by the rays being split
into distinct bundles
as they propagate through the object.
This splitting is due
to the discontinuous gradient field
inherit in trilinear interpolation.
As the resolution of the voxelizations increase
in the second row of the figure,
the resulting images progress
through a series of Moiré patterns
as the artifacts become smaller than the pixel size.
At high enough voxel-to-pixel ratio,
these artifacts should disappear
as they get averaged over the surface of the pixel,
but they never disappear in an absolute sense.
The tricubic interpolation
also suffers from artifacts
at low voxel resolutions,
due to the discrepancy
between the voxelized object and true object,
but the resulting image
is always continuous and the artifacts
disappear in an absolute sense
at high voxel resolutions.

These results suggest that tricubic interpolation
should be preferred over trilinear interpolation,
though it should be mentioned that trilinear interpolation,
being first-order in the sense that,
to perform a single interpolation,
one needs the 8 neighboring voxel values.
Tricubic is second-order in that it also requires
the neighbors' neighbors,
thus 64 total voxel values.
The linear technique thus requires
approximately eight-times less computation.
A compromise may be to use a different linear technique
that has a continuous second derivative,
such as Hanning interpolation
\cite{denis1995ultrasonic},
to maintain the efficiency
while avoiding the particular artifacts
caused by trilinear interpolation.
Investigating this technique
is left for future work.

For an appropriately band-limited object,
all of these interpolation techniques
are effectively approximations
of Shannon interpolation,
which can reproduce the underlying object exactly
from the voxel values.
Shannon sampling is
the highest-possible-order interpolation,
that is, that every voxel value
influences every interpolation.
It thus is very slow,
but the Shannon perspective is helpful
to glean a heuristic from our results.

The Shannon criterion for perfect reconstruction
is that the sampling frequency
is more than twice the highest frequency
present in the underlying object.
If an object has
any significant spatial frequency components
beyond the cutoff implied by a chosen voxel size,
its voxelization will have aliasing artifacts
that will in turn cause artifacts in the refraction.
The Gaussian object used
in section \ref{section_gaussian}
is not strictly band-limited,
but has vanishingly small high-frequency components
and, thus, vanishingly small aliasing artifacts.
We can decide at what point
we consider them negligible.
If we define the error between
the true Gaussian, $g$,
and the voxelized Gaussian
interpolated with Shannon interpolation,
$g'$,
as
	\begin{equation}
	\text{error} = \frac{
			\iiint d^3r \; |g(\vec{r} \, ) - g'(\vec{r} \, )|^2
		}{
			\iiint d^3r \; |g(\vec{r} \, )|^2
		}
	\end{equation}
this error falls below 0.1\%
when the ratio of the Gaussian's width to voxel size
is greater than 2.49. 
We'll take this to define
an effective band limit of the Gaussian.
If we consider the result in
Fig.~\ref{fig_voxel_gaussian}r,
which has a width-to-voxel-size ratio of 16,
as satisfactory,
this suggests that interpolating
at about 6.4 times the Shannon criterion
is sufficient for simulating
with tricubic interpolation,
that is,
the sampling frequency
is more than 12.8 the highest frequency
present in the underlying object.
This frequency cutoff can be enforced
by applying an appropriate anti-aliasing filter
prior to voxelization.

CT reconstructions
are naturally blurred at some scale
due to the point-spread function
of the imaging apparatus,
and are thus band-limited in spatial frequency.
The magnification of the imaging setup
is thus typically adjusted
such that the pixel size
samples the tomograms
just above the Shannon criterion.
The resulting reconstruction
will have a voxel size
that satisfies the Shannon criterion
for a band-limited version
of the underlying 3D object.
The heuristic recommended here is to use voxels
that are several times finer
than the Shannon criterion to
compensate for the algorithm
using tricubic interpolation
rather than Shannon interpolation.
To achieve this from a given CT voxel volume,
the volume must be upsampled
using Shannon interpolation
by the appropriate factor
before being input into the algorithm.

The results in
Fig.~\ref{rotation_fig}
demonstrate that the artifacts
from the voxel-grid-aligned illumination
present in Fig.~\ref{fig_voxel_gaussian}
largely disappear when the illumination
is misaligned from the voxel grid.
Even the voxelization that gave the worst result
in the previous experiment
yields an image that is visually indistinguishable
from the ground truth
when illuminated far from alignment.
This suggests that our heuristic
can be less strict
for well-misaligned illumination
by approximately a factor of two.

The images in section \ref{section_lumpy}
illustrate the type of signal one can
expect to see when X-ray light
refracts in a material
with significant random variation
at some length scale,
as is common in biological samples like bone.
The object with coarser features
is voxelized with a sampling frequency
to band-limit ratio of about 12,
just below our recommendation
for voxel-aligned illumination.
The object with finer features 
has a sampling-frequency-to-band-limit
ratio of about 9.

In Fig.~\ref{fig_lumpy_histogram_compound},
where the objects are illuminated with voxel-aligned rays,
the image from
the trilinearly interpolated coarser object
has the same
blocky, discontinuous artifacts as before,
and the image
using tricubic interpolation
has no visible artifacts, as expected.
In the images from the finer object,
the trilinear artifacts are worse,
and there are slight voxelization artifacts
visible in the tricubic image as well.
In both cases,
the tricubic interpolation
images are clearly superior.
In Fig.~\ref{fig_lumpy_histogram_compound_rot},
where the objects are illuminated with rays
that are well-misaligned from the voxel grid,
the images using trilinear and tricubic interpolation
are visually indistinguishable.
From the difference plots,
we can see some caustics
are shifted by a single pixel.
This suggests that using
trilinear interpolation
can produce acceptable simulations
in the case where the rays never reach
alignment with the voxel grid.

This technique has shown initial success
for simulating X-ray images in the case
that the object is purely refractive.
It can be readily adapted to include
incoherent scattering
and absorption.
This will allow for detailed simulations
of X-ray dose in future work.

Importantly, this method as presented
cannot simulate coherence effects
which are crucial for some imaging techniques.
In future work,
it can be adapted analogously
to previous techniques developed using straight
rays that refract only at sharp interfaces
\cite{peter2014combining}.
In this strategy, each ray carries a phase
determined by a corresponding path integral
through the index-of-refraction volume.
The rays incident on a pixel
are then added
according to their phases
to reconstruct the incident field.
As this technique is calculating the field
with what is essentially
a Monte Carlo implementation of a Feynman path integral
rather than the intensity
with a classical Monte Carlo integral,
there are additional complexities to consider
in representing a given illumination,
and the technique is computationally expensive.
An alternative strategy would be to adapt
a technique developed
to simulate visible light in computer graphics
known as wave-tracing
which avoids some of the complexities
by propagating Gaussian beams rather than rays
\cite{steinberg2025wave}.




\section{Conclusion}
In this work we have demonstrated a 
first-order ray-tracing method
to simulate X-ray images
directly on voxel grids.
Validating this method
compared to the well-known analytic solution
of the brachistochrone problem
motivated a heuristic
 to determine a sufficient step size
based on the maximum gradient
of the index of refraction
in a voxelized object.
In investigating
how the propagation method interacts
with voxel representations
of interfaces,
we found that sharp interfaces
can be represented
with smooth transitions
at the expense of artifacts
at the scale of the width of the transition.
In our numerical experiments using the method in 3D,
we found that the trilinear interpolation introduces
detrimental artifacts due
to the discontinuous gradient field.
Tricubic interpolation remedies this particular problem,
but is still susceptible to artifacts at low voxel resolutions.
We found that voxelizations at least 6.4 times the
Shannon criterion are necessary to produce quality results.
We also observed that the voxelization artifacts
are significantly reduced
in case of plane-wave illumination
that is far from aligned with the voxel grid.

\appendix
\section{Derivation}
\label{appendix_derivation}
\subsection{Eikonal Equation}
Maxwell's equations written in Lorentz-Heaviside units
in linear isotropic media with no free changes
and no constant components of the field are
	\begin{equation*}
	\vec{\nabla} \times \vec{E}
	= -\frac{\mu}{c} \, \frac{\partial}{\partial t} \vec{H}
	\label{curl_E}
	\end{equation*}
	\begin{equation}
	\vec{\nabla} \times \vec{H}
	= \frac{\varepsilon}{c} \, \frac{\partial}{\partial t} \vec{E}
	\; ,
	\label{curl_H}
	\end{equation}
where $\vec{E}$ and $\vec{H}$ are the electric and magnetic fields
and $\epsilon$ and $\mu$ are the permittivity and permeability.
If $\vec{E}$ and $\vec{H}$ are written
in terms of their spatial part and frequency components,
	\begin{equation*}
	\vec{E}(\vec{r}, t)
	= \int \frac{d\omega}{2 \pi} \,
	e^{i \omega t} \, \mathcal{F}_t[\vec{E}](\vec{r}, \, \omega) 
	\label{fourier_E}
	\end{equation*}
	\begin{equation}
	\vec{H}(\vec{r}, t)
	= \int \frac{d\omega}{2 \pi} \,
	e^{i \omega t} \, \mathcal{F}_t[\vec{H}](\vec{r}, \, \omega) 
	\; ,
	\label{fourier_H}
	\end{equation}
where $\mathcal{F}_t$ denotes a Fourier transform
with respect to time,
Eqs.~\ref{curl_H} become
	\begin{equation*}
	\vec{\nabla} \times \mathcal{F}_t[\vec{E}]
	= -\frac{i \omega \mu}{c} \, \mathcal{F}_t[\vec{H}]
	\label{curl_E_fourier}
	\end{equation*}
	\begin{equation}
	\vec{\nabla} \times \mathcal{F}_t[\vec{H}]
	= \frac{i \omega \varepsilon}{c} \, \mathcal{F}_t[\vec{E}]
	\label{curl_H_fourier}
	\end{equation}
for each $\omega$.
We can write
$\mathcal{F}_t[\vec{E}]$ and 
$\mathcal{F}_t[\vec{H}]$ as
real amplitude functions multiplied by a phase function.
	\begin{equation*}
	\mathcal{F}_t[\vec{E}]
	= \vec{A}_E(\vec{r} \,) \, e^{i \phi(\vec{r} \,)}
	\label{phase_amp_E}
	\end{equation*}
	\begin{equation}
	\mathcal{F}_t[\vec{H}]
	= \vec{A}_H(\vec{r} \,) \, e^{i \phi(\vec{r} \,)}
	\label{phase_amp_H}
	\end{equation}
Substituting these into
Eqs.~\ref{curl_H_fourier}
yields
	\begin{equation*}
	\vec{\nabla} \times \vec{A}_E
	+ i \, \vec{\nabla} \phi \times \vec{A}_E
	= - \frac{i \omega \mu}{c} \vec{A}_H
	\end{equation*}
	\begin{equation}
	\vec{\nabla} \times \vec{A}_H
	+ i \, \vec{\nabla} \phi \times \vec{A}_H
	= \frac{i \omega \varepsilon}{c} \vec{A}_E
	\; .
	\end{equation}
Here we will take the limit
that $\vec{A}_E(\vec{r} \,)$ and $\vec{A}_H(\vec{r} \,)$
vary much more slowly than the phase,
thus the curl of the amplitude functions
are negligible in the above equation,
leaving
	\begin{equation*}
	\vec{\nabla} \phi \times \vec{A}_E
	= - \frac{\omega \mu}{c} \vec{A}_H
	\label{curl_approx_E}
	\end{equation*}
	\begin{equation}
	\vec{\nabla} \phi \times \vec{A}_H
	= \frac{\omega \varepsilon}{c} \vec{A}_E
	\; .
	\label{curl_approx_H}
	\end{equation}
Eqs.~\ref{curl_approx_H}
together imply
	\begin{equation}
	\vec{A}_E \, \bcdot \, \vec{A}_H = 0
	\label{A_orthogonal}
	\end{equation}
	\begin{equation}
	\hat{\nabla} \phi
	= \frac{\vec{A}_E \times \vec{A}_H}{|\vec{A}_E \times \vec{A}_H|}
	\label{AXA}
	\end{equation}
	\begin{equation}
	|\vec{\nabla} \phi|^2 = \frac{\omega^2 \mu \varepsilon}{c^2}
	\end{equation}
	\begin{equation}
	A_E = \sqrt{\frac{\mu}{\varepsilon}} \, A_H
	\; .
	\end{equation}
	Let's define
	\begin{equation}
	k^2(\vec{r} \, )
	= \frac{\omega^2 \, \mu(\vec{r} \, ) \, \varepsilon(\vec{r} \, )}{c^2}
	\; .
	\end{equation}
We thus have what is called the eikonal equation.
	\begin{equation}
	|\vec{\nabla} \phi(\vec{r} \, )| = k(\vec{r} \, )
	\label{eikonal_equation}
	\end{equation}

\subsection{Ray Equation}
\label{appendix_ray_equation}
Rays are defined as paths
that are tangent to $\vec{\nabla} \phi$,
that is,
	\begin{equation}
	\frac{d}{d\tau}[\vec{r} \, ](\tau)
	= \eta(\vec{r}(\tau)) \, \vec{\nabla} \phi(\vec{r}(\tau))
	\; ,
	\end{equation}
where $\eta$
defines our choice of parameterization
for the path.
Using the eikonal equation,
Eq.~\ref{eikonal_equation},
this is equivalent to
	\begin{equation}
	\frac{1}{\eta} \,
	\frac{d}{d\tau} \left[ \frac{1}{\eta} \, \frac{d\vec{r}}{d\tau} \right]
	= \frac{1}{2} \, \vec{\nabla} \left[ k^2 \right]
	\; ,
	\label{general_wave_equation}
	\end{equation}
which we will call the general ray equation.
The eikonal equation and general ray equation,
Eq.~\ref{eikonal_equation} and \ref{general_wave_equation},
together imply that phase accumulated along a ray is
	\begin{equation}
	\phi(\vec{r}(\tau_1)) - \phi(\vec{r}(\tau_0))
	= \int_{\tau_0}^{\tau_1} d\tau \,
	\eta(\vec{r}(\tau)) \, k^2(\vec{r}(\tau))
	\; .
	\end{equation}
The general ray equation,
also implies a conserved quantity along rays,
that is,
	\begin{equation}
	\frac{1}{\eta^2} \left| \frac{d \vec{r}}{d \tau} \right| ^2
	- k^2
	= \text{constant}
	\; .
	\end{equation}
We will only consider rays
in which this constant is zero,
that is,
	\begin{equation}
	\left| \frac{d \vec{r}}{d \tau} \right|
	= \frac{k}{\eta}  
	\; ,
	\end{equation}
thus determining what we'll call the ray speed.

There are a few choices of $\eta$ that are interesting.
Choosing $\eta = 1$
results in making the ray equation
analogous to a Newtonian mechanics case
of a point particle
moving under the influence of a potential.
Choosing $\eta = \frac{1}{k^2}$ results
in the parameter $\tau$
being equal to the accumulated phase
and the ray speed speed being proportional
to the speed of light in the medium.
Choosing $\eta = \frac{1}{k}$ results
in the ray speed being 1,
and the parameter $\tau$
being equal to the arclength of the ray path.
This last option is the choice we will focus on.
With this choice the ray equation becomes
	\begin{equation}
	\frac{d}{d\tau} \left[ k \, \frac{d\vec{r}}{d\tau} \right]
	= \vec{\nabla} k
	\; ,
	\end{equation}
which can be expanded as
	\begin{equation}
	\vec{\nabla} k \, \bcdot \, \frac{d\vec{r}}{d\tau}
	\, \frac{d\vec{r}}{d\tau}
	+ k \, \frac{d^2 \vec{r}}{d \tau^2}
	- \vec{\nabla} k
	= 0
	\; .
	\label{ray_equation}
	\end{equation}
This is the version we refer to simply as the ray equation.
The accumulated phases are calculated as
	\begin{equation}
	\phi(\vec{r}(\tau_1)) - \phi(\vec{r}(\tau_0))
	= \int_{\tau_0}^{\tau_1} d\tau \,
	k(\vec{r}(\tau))
	\end{equation}

\subsection{Amplitude Calculation}
\label{appendix_amplitude}
For any choice of parameterization,
the amplitudes,
$A_E$ and $A_H$,
can be calculated
by considering the Poynting vector,
the vector describing
the energy flux density
of our electromagnetic field,
and boundary conditions.
The Poynting vector is
	\begin{equation}
	\vec{S} = \vec{E} \times \vec{H}
	\; ,
	\end{equation}
and obeys Poynting's theorem, that is
	\begin{equation}
	\vec{\nabla} \, \bcdot \, \vec{S}
	= - \frac{\partial u}{\partial t}
	\label{poynting_theorem}
	\end{equation}
in the absence of free currents,
where u is the energy density
	\begin{equation}
	u = \frac{1}{2}(\varepsilon \, E^2 + \mu \, H^2)
	\; .
	\end{equation}

We can use the Poynting vector
to model a measurement.
Lets assume a detector
measures the amount of energy
that is deposited on it over time.
Assuming perfect absorption efficiency,
this quantity can be calculated as
	\begin{equation}
	I_\text{surface} = \int_{-\infty}^\infty dt \, \iint_\text{surface} d\vec{a} \, \bcdot \, \vec{S}
	\end{equation}
Where here we are assuming a normalizable pulse.

Using Eqs.~\ref{fourier_H},
\ref{phase_amp_H},
\ref{AXA}, \ref{eikonal_equation}
and that the field values are real,
we can write
	\begin{equation}
		\begin{aligned}
		I_\text{surface} &= \int \frac{d\omega}{2 \pi} \,
		\iint_\text{surface} d\vec{a} \, \bcdot \,
		\mathcal{F}_t[\vec{E}]
		\times \mathcal{F}_t^*[\vec{H}] \\
		&= \int \frac{d\omega}{2 \pi} \,
		\iint_\text{surface} d\vec{a} \, \bcdot \, \hat{k} \,
		A_E \, A_H
		\end{aligned}
	\end{equation}
We'll consider
	\begin{equation}
	\mathcal{F}_t[I_\text{surface}]
	= \iint_\text{surface} d\vec{a} \, \bcdot \, \hat{k} \,
	A_E \, A_H
	\end{equation}
which is proportional
to the amount of energy
per frequency in our pulse.

We can calculate a similar integral
over a finite closed surface
	\begin{equation}
		\begin{aligned}
		I_\text{closed} &= \int_{-\infty}^{\infty} dt \,
		\iint_\text{closed} d\vec{a} \, \bcdot \, \vec{S} \\
		&= \int_{-\infty}^{\infty} dt \,
		\iiint_\text{volume} d^3 r \,
		\frac{\partial u}{\partial t} \\
		&= 0
		\end{aligned}
	\end{equation}
using Eq.~\ref{poynting_theorem}
and the divergence theorem.
This quantity is zero
because the total energy
enclosed by the surface
at infinite time in the past an in the future
must be zero for a normalizable pulse
evolving according to Maxwell's equations.
A periodic incident field
works out analogously
with a average energy per period
in place of the total incident energy
over all time.

Lets consider the intensity
integrated over a special closed surface
that we will call a ray tube.
These ray tubes
are defined as having two end surfaces
connected by walls parallel
to the ray flow defined in
Eq.~\ref{ray_equation}.
This means if one of the end surfaces
is defined,
the boundary of the other end surface
is constrained.
These two end surfaces do not need to be flat;
they can bow in our out arbitrarily.
Integrating over the surface of a ray tube gives
	\begin{equation}
    \begin{aligned}
		\mathcal{F}_t[I_\text{ray tube}]
		&= \iint_\text{end 1} d\vec{a}
		\, \bcdot \, \hat{k} \,
		A_E \, A_H \\
		&- \iint_\text{end 2} d\vec{a}
		\bcdot \, \hat{k} \,
		A_E \, A_H \\
		&+ \iint_\text{walls} d\vec{a}
		\, \bcdot \, \hat{k} \,
		A_E \, A_H \\
		&= 0
    \end{aligned}
	\end{equation}
where the minus sign
on the middle integral
indicates we are defining
the normal vector
of the beam surface as inwards.
By construction,
the normal vector of the walls
is orthogonal to the ray direction,
thus the walls integral is zero. We thus have,
	\begin{equation}
	\mathcal{F}_t[I_\text{end 1}]
	= \mathcal{F}_t[I_\text{end 2}]
	\end{equation}
that is, energy carried
in each frequency component
is conserved along ray tubes.

Let us assume
we are illuminating an object
described by $k(\vec{r} \,)$
with light from a beam
whose field
is known exactly
along the plane at $z=0$,
and whose ray direction, $\hat{k}$,
is in the $\hat{z}$ direction
everywhere along this plane.
Let us consider a pixel
of an imaginary detector
in the plane $z = z_d$.
We can imagine a ray tube
that connects that pixel
to the known boundary on the beam plane.
The ray tube has one simple
and one complicated surface.
One surface is simply the surface of
the pixel in the detector plane
and the other
that is some distorted shape
in the beam plane at $z=0$.
	\begin{equation}
	  \begin{aligned}
		\mathcal{F}_t[I_\text{pixel}]
		&= \iint_\text{pixel} d^2 r_\perp \,
		\hat{z} \bcdot \hat{k}(\vec{r}_\perp, \, z_d) \,
		A_E(\vec{r}_\perp, \, z_d) \,
		A_H(\vec{r}_\perp, \, z_d) \\
		&= \iint_\text{distorted shape} d^2 r_\perp' \,
		A_E(\pvec{r}_\perp, \, 0) \,
		A_H(\pvec{r}_\perp, \, 0)
	  \end{aligned}
	\end{equation}
Let us define a function
	\begin{equation}
	\Theta_\text{pixel}(\vec{r}_\perp)
	=
		\begin{cases}
		1 & \text{for } \vec{r}_\perp \text{ inside pixel}
		\\
		0 & \text{for } \vec{r}_\perp \text{ outside pixel}
		\end{cases}
	\end{equation}
We can thus write
	\begin{multline}
	\mathcal{F}_t[I_\text{pixel}] = \\
	\iint d^2 r_\perp \, 
    \hat{z} \bcdot \hat{k}(\vec{r}_\perp, \, z_d) \,
	\Theta_\text{pixel}(\vec{r}_\perp) \,
	A_E(\vec{r}_\perp, \, z_d) \,
	A_H(\vec{r}_\perp, \, z_d)
	\end{multline}
where our bounds now cover the entire detector plane.
Similarly, we can define a function
for the distorted shape project
on the beam plane by the ray equation
	\begin{equation}
		\begin{aligned}
		\Theta_\text{distorted shape}(\pvec{r}_\perp)
		&=
			\begin{cases}
			1 & \text{for } \vec{r}_\perp(\pvec{r}_\perp)
			\text{ inside pixel}
			\\
			0 & \text{for } \vec{r}_\perp(\pvec{r}_\perp)
			\text{ outside pixel}
			\end{cases}
		\\
		&= \Theta_\text{pixel}(\vec{r}_\perp(\pvec{r}_\perp))
		\end{aligned}
	\end{equation}
where $\vec{r}_\perp(\pvec{r}_\perp)$
is the the result of ray tracing
from the point $\pvec{r}_\perp$
in the beam plane to the detector plane.
We can write
	\begin{equation}
	\mathcal{F}_t[I_\text{pixel}]
	= \iint d^2 r_\perp' \, 
	\Theta_\text{pixel}(\vec{r}_\perp(\pvec{r}_\perp)) \,
	A_E(\pvec{r}_\perp, \, 0) \,
	A_H(\pvec{r}_\perp, \, 0)
	\end{equation}
Lets define
	\begin{equation}
	I_0 \, p(\pvec{r}_\perp)
	= A_E(\pvec{r}_\perp, \, 0) \,
	A_H(\pvec{r}_\perp, \, 0)
	\end{equation}
where $p$ is a probability distribution
over the beam plane, that is 
	\begin{equation}
	\iint d^2 r_\perp' \, 
	p(\pvec{r}_\perp) 
	= 1
	\end{equation}
We can thus write our pixel intensity integral as
	\begin{equation}
	\begin{aligned}
	\mathcal{F}_t[I_\text{pixel}]
	 & = I_0 \, \iint d^2 r_\perp' \, 
	p(\pvec{r}_\perp) \,
	\Theta_\text{pixel}(\vec{r}_\perp(\pvec{r}_\perp)) \\
	& = I_0 \, \sum_n^N
	\Theta_\text{pixel}(\vec{r}_\perp(\pvec{r}_{\perp n})) \,
    \end{aligned}
    \end{equation}
where $\pvec{r}_{\perp n}$
is the $n^\text{th}$ sample of the distribution $p$.
Assuming monochromatic plane-wave illumination,
the procedure to simulate images is thus
to first sample initial ray positions from the beam,
to propagate the rays to the detector according to the ray equation,
then to count the rays that intersect each pixel.

\section{Interpolation}
Given a scalar function
defined on a voxel grid,
interpolation extends
this function to be defined
on all intermediate values
between grid points.
This extended function
also defines the gradient.
There are many types of interpolation.
Here we will elaborate on two.

\subsection{Trilinear Interpolation}
\label{trilinear_appendix}
Given the values, $f_{lmn}$,
and grid positions, $\vec{r}_{lnm}$,
of a voxel function
we wish to interpolate,
and the position, $\vec{r}$,
at which we wish to interpolate it,
we can calculate
a trilinear interpolation
and its gradient
with the following procedure.

First we identify the smallest indices
that correspond to a grid point
adjacent to $\vec{r}$.
That is
	\begin{equation*}
	(x - x_0) \, // \, \Delta r = l_c
	\end{equation*}
	\begin{equation*}
	\; \;
	(y - y_0) \, // \, \Delta r = m_c
	\end{equation*}
	\begin{equation}
	\; \; \;
	(z - z_0) \, // \, \Delta r = n_c
	\; ,
	\label{reference_neighbor_indices}
	\end{equation}
where $x_0$, $y_0$, and $z_0$
are the position coordinates
of the $000^\text{th}$ grid point,
$\Delta r$ is the grid spacing,
and $//$ represents floor division.
With these,
we can find the values
of the 8 adjacent voxels.
We'll denote these 8 as
	\begin{equation}
	f'_{l'm'n'} = f_{\, l_c + l', \,  m_c + m', \, n_c + n'}
	\label{f_neighbors}
	\end{equation}
where $l'$, $m'$, and $n'$
each range from 0 to 1.

Second, if we define
the position in question relative
to this adjacent grid point
in voxel length units,
	\begin{equation}
		\begin{bmatrix} 
		x' \\
		y' \\
		z' \\
		\end{bmatrix}
	= \frac{\vec{r} - \vec{r}_{l_c m_c n_c}}{\Delta r}
	\; ,
	\label{xyz}
	\end{equation}
and define the matrix
	\begin{equation}
	M =
		\begin{bmatrix} 
		1 & -1 \\
		0 & 1 \\
		\end{bmatrix}
	\; ,
	\end{equation}
we can construct the vectors
	\begin{equation}
	\vec{X} =
		\begin{bmatrix} 
		1 \\
		x' \\
		\end{bmatrix} \; ,
	\; \; \;
	\vec{Y} =
		\begin{bmatrix} 
		1 \\
		y' \\
		\end{bmatrix}
	\; \; \text{and} \; \;
	\vec{Z} =
		\begin{bmatrix} 
		1 \\
		z' \\
		\end{bmatrix}
	\; ,
	\end{equation}
and calculate the interpolation as 
	\begin{multline}
	f(\vec{r} \, ) =\\
	\sum_{i'j'k'l'm'n'} f_{l'm'n'} \, M_{i'l'} \, M_{j'm'} \, M_{k'n'} \, X_{i'} \, Y_{j'} \, Z_{k'}
	\; .
	\label{interpolate_sum}
	\end{multline}
The gradient is
	\begin{multline}
	\vec{\nabla} f(\vec{r} \, ) =
	\frac{1}{\Delta r} \\
	\sum_{i'j'k'l'm'n'} \, f_{l'm'n'} \, M_{i'l'} \, M_{j'm'} \, M_{k'n'}
		\begin{bmatrix} 
		\delta_{x'} X_{i'} \, Y_{j'} \, Z_{k'} \\
		X_{i'} \, \delta_{y'} Y_{j'} \, Z_{k'} \\
		X_{i'} \, Y_{j'} \, \delta_{z'} Z_{k'} \\
		\end{bmatrix}
	\; ,
	\label{grad_interpolate_sum}
	\end{multline}
where
	\begin{equation}
	\delta_{x'} \vec{X} =
	\delta_{y'} \vec{Y} =
	\delta_{z'} \vec{Z} =
		\begin{bmatrix} 
		0 \\
		1 \\
		\end{bmatrix}
	\; .
	\end{equation}

\subsection{Tricubic Interpolation}
\label{tricubic_appendix}
For tricubic interpolation,
the process is analogous to section
\ref{trilinear_appendix},
with the key difference
that we must consider
not just the 8 neighboring voxels
adjacent to the interpolation position,
but also the neighbors' neighbors,
bringing the total to 64 voxels
in consideration.
With the indices
of the adjacent reference voxel
$l_c$, $m_c$, and $n_c$
calculated as in
Eq.~\ref{reference_neighbor_indices},
we'll denote the voxel values
considered in our calculation,
$f_{l'm'n'}$,
the same way as in Eq.~\ref{f_neighbors},
but with the $l'$, $m'$, and $n'$
indices running from -1 to 2.

The interpolation calculation
takes the same form
as Eq.~\ref{interpolate_sum}
but with the matrix $M$
defined as
	\begin{equation}
	M =
		\begin{bmatrix} 
		0   & 1    & 0    & 0  \\
		a   & 0    & -a   & 0  \\
		-2a & -a-3 & 2a+3 & a  \\
		a   & a+3  & -a-2 & -a \\
		\end{bmatrix}
	\; ,
	\end{equation}
where we take $a=-0.5$,
and the vectors
$\vec{X}$,
$\vec{Y}$,
and $\vec{Z}$
defined as
	\begin{equation}
	\vec{X} =
		\begin{bmatrix} 
		1 \\
		x' \\
		{x'}^2 \\
		{x'}^3 \\
		\end{bmatrix} \; ,
	\; \; \;
	\vec{Y} =
		\begin{bmatrix} 
		1 \\
		y' \\
		{y'}^2 \\
		{y'}^3 \\
		\end{bmatrix}
	\; \; \text{and} \; \;
	\vec{Z} =
		\begin{bmatrix} 
		1 \\
		z' \\
		{z'}^2 \\
		{z'}^3 \\
		\end{bmatrix}
	\; ,
	\end{equation}
where $x'$, $y'$, and $z'$
are defined
as in Eq.~\ref{xyz}.
We've written the matrix and vectors
such that values
with a $-1$ index
are in the top row
or furthest left column.
The gradient calculation
has the same form as
Eq.~\ref{grad_interpolate_sum}
with
	\begin{equation}
	\delta_{x'} \vec{X} =
		\begin{bmatrix} 
		0 \\
		1 \\
		x' \\
		{x'}^2 \\
		\end{bmatrix} \; ,
	\; \; \;
	\delta_{y'} \vec{Y} =
		\begin{bmatrix} 
		0 \\
		1 \\
		y' \\
		{y'}^2 \\
		\end{bmatrix}
	\; \; \text{and} \; \;
	\delta_{z'} \vec{Z} =
		\begin{bmatrix} 
		0 \\
		1 \\
		z' \\
		{z'}^2 \\
		\end{bmatrix}
	\; .
	\end{equation}

\section{Path Error}
\label{path_error_section}
We define a path error metric
to compare two paths
of the same arclength
as
	\begin{equation}
	\text{path error}
	= L^{-\frac{3}{2}}
	\sqrt{\int_0^L ds \,
	\left| \vec{r}(s)
	- {\vec{r}} \, '(s) \right|^2}
	\; ,
	\label{path_error}
	\end{equation}
where $s$
is the accumulated arclength parameter,
$\vec{r}$ and $\vec{r} \, '$
are the two paths,
and $L$ is the arclength
of the paths.
Comparing a path with itself
displaced by its own path length
yields a path error of 1.

\section{Image root mean square error}
\label{rms_error_section}
To compare
a simulated image
in question
with an image
taken to be the ground truth,
we define an image error metric
as
	\begin{equation}
	\text{image error}
	= \frac{\sqrt{\sum_{ij} (A_{ij} - B_{ij})^2}}
	{\sqrt{\sum_{ij} B_{ij}^2}}
	\; ,
	\label{rms_error}
	\end{equation}
where $A$ is the image in question,
and $B$ is the ground truth.
Comparing an image
consisting only of zeros
with any ground truth
yields an error of 1.


\bibliography{ref.bib}
\end{document}